\documentstyle[aps,preprint,graphicx]{revtex}
\tighten
\draft

\begin{document}

\title{Radiative corrections in neutrino-deuterium disintegration\\}
\author {A. Kurylov$^{a,b}$, M.J. Ramsey-Musolf$^{a,b}$, and P. Vogel$^{a}$ }
\address{$^{a}$Kellogg Radiation Laboratory and Physics Department \\
Caltech, Pasadena, CA 91125 \\
$^b$ Department of Physics, University of Connecticut, Storrs, CT  06269}
\date{\today}
\maketitle

\begin{abstract}
The radiative corrections of order $\alpha$ for the charged and neutral
current  neutrino-deuterium disintegration for energies relevant to the SNO
experiment are evaluated. Particular attention is paid to the issue
of the bremsstrahlung detection threshold. It is shown that the radiative
corrections to the total cross section for the charged current reaction
are independent of that threshold, as they must be for consistency, and
amount to a slowly decreasing function of the neutrino energy $E_{\nu}$,
varying from $\sim$ 4\% at low energies to $\sim$ 3\% at the end of the
$^8$B spectrum. The differential cross section corrections, 
on the other hand, do depend
on the  bremsstrahlung detection threshold. Various choices of the
threshold are discussed. It is shown that for a realistic choice
of the threshold, and for the actual electron energy threshold
of the SNO detector, the deduced $^8$B $\nu_e$ flux should be decreased
by $\sim$ 2\%. The radiative corrections to the neutral current reaction
are also evaluated. \\ \\
\end{abstract}

PACS number(s): 13.10.+q, 13.15.+g, 25.30.Pt

\vspace{1cm}

\section{Introduction}

Solar neutrinos from $^8$B decay have been detected at
the Sudbury Neutrino Observatory (SNO)  \cite{SNO}
via the charged current (CC) reaction
\begin{equation}
\label{eq:ccreaction}
\nu_e + d \rightarrow p + p + e^- ~.
\label{eq:CC}
\end{equation}
In the next phase of the SNO experiment, currently underway, the rate of
the neutral current deuteron disintegration (NC)
\begin{equation}
\label{eq:ncreaction}
\nu_e + d \rightarrow p + n + \nu_e ~
\label{eq:NC}
\end{equation}
will be also measured.

From the measurement of the CC reaction rate
the flux at Earth of the $^8$B solar $\nu_e$
was determined to be \cite{SNO},
\begin{equation}
\Phi_{SNO}^{CC}(\nu_e) =
1.75 \pm 0.07({\rm stat})_{-0.11}^{+0.12}({\rm syst})
\pm 0.05({\rm theor}) \times 10^6 {\rm cm}^{-2}{\rm s}^{-1}~.
\label{fl:CC}
\end{equation}
The $^8$B solar neutrinos were also detected in the precision
measurement by the Super-Kamiokande Collaboration
(SK) \cite{SK} using the elastic
scattering (ES) on electrons. That reaction is sensitive not only to
the charged current weak interaction but also to the neutral current
interaction. From the SK measurement
the  $^8$B solar neutrino flux was deduced to be
\begin{equation}
\label{fl:NC}
\Phi_{SK}^{ES}(\nu_e) = 2.32 \pm 0.03({\rm stat})_{-0.07}^{+0.08}({\rm syst})
\times 10^6 {\rm cm}^{-2}{\rm s}^{-1}~.
\end{equation}
The difference between these two flux determinations, at the 3.3
$\sigma$ level, can be regarded as a `smoking gun' proof of neutrino
oscillations, independent of the
solar model flux calculation. By itself, $\nu_e e$ NC scattering cannot account
for the difference between (\ref{fl:CC}) and (\ref{fl:NC}). The excess ES
events
must involve a neutrino species which contributes disproportionately to the
NC rate.
According to the
oscillation hypothesis, some of the  $^8$B solar $\nu_e$ oscillate
into another
active neutrino flavor $\nu_{\mu \tau}$.
These $\nu_{\mu \tau}$ neutrinos then cannot cause the charged
current reaction, Eq.(\ref{eq:CC}), but they can and do undergo NC
scattering on
electrons. Assuming that this is what is really happening, one
arrives at the total $^8$B solar neutrino flux consistent
with the Standard Solar Model \cite{BP,TC}.
This agreement may be used as a  supporting
evidence for the oscillation hypothesis, which will be further tested by
comparing
the CC and NC reaction rates
measured by the SNO experiment alone.

The goal of the present work is the evaluation of the radiative
(order $\alpha$) corrections to the cross sections of the CC and NC
reactions. Precise knowledge of these cross sections
has obvious relevance for the determination of  the  $^8$B neutrino
flux. Experimentally, one measures the number and energies
of the electron events for the CC reaction,
or the number of neutron events for the NC reaction, which
after corrections for cuts and experimental efficiencies, is
an integral over the incoming neutrino energies
of the  $^8$B solar neutrino flux (possibly modified by the
neutrino oscillations) times the differential cross section.
Hence any error in the cross section causes a corresponding error
in the deduced flux.

In analyzing the SNO CC data the theoretical cross section of
Ref. \cite{cross} was used. The assumed uncertainty of the
calculated cross section is reflected in the theoretical
uncertainty of the deduced flux, Eq.(\ref{fl:CC}). However,
radiative corrections were not applied to the CC cross section.

The radiative corrections to the CC reaction (\ref{eq:CC}) were
evaluated by Towner \cite{Towner}. 
That analysis was recently questioned by Beacom and
Parke\cite{John}, who noted that the total CC cross section for detected and
undetected bremsstrahlung differ, according to the analysis of Ref.
\cite{Towner}.
Such a difference is unphysical. The observation of Ref. \cite{John} has
lead to
understandable confusion among experimentalists as to the appropriate radiative
corrections to apply to the SNO data. While the published SNO result did
not include
any radiative corrections, the level of confidence in future CC/NC
comparisons could
depend significantly on a proper treatment of the radiative corrections.
Thus, in
what follows we revisit the analysis of Ref. \cite{Towner}, in an effort to
resolve the present controversy.

While we have no quarrel with the basic treatment of the radiative corrections
to the CC reaction in  \cite{Towner}, we confirm the observations of Ref.
\cite{John} and identify the origins of the inconsistency in Towner's results:
(a) neglect of a strong momentum-dependence in the Gamow-Teller
$^3$S$_1\to^1$S$_0$
matrix element, and (b) improper ordering of limits involving
$E_\gamma^{min}$ and the infrared regulator. After correcting for these
issues, we obtain identical total
CC cross sections for detected and undetected
bremsstrahlung. The results imply an $E_\nu$-dependent correction to the total
CC cross section which varies from $\sim 4\%$ to $\sim 3\%$ over the range of
available neutrino energies.

In addition to foregoing, we also recast the treatment in Ref. \cite{Towner} of
hadronic effects in the radiative corrections into the language of
effective field
theory (EFT). Although the traditional treatment in Ref. \cite{Sirlin} and EFT
frameworks are equivalent, the latter provides a systematic approach for
long-distance, hadronic effects presently uncalculable from first
principles in QCD.
As discussed in Ref. \cite{Sirlin}, matching the asymptotic and long-distance
calculations (in EFT) involves use of a hadronic scale,
$M_{had}$,  whose
choice introduces a  small theoretical uncertainty into the radiative
corrections.
We argue that the choice of $M_{had}$ made in Ref.
\cite{Towner} is
possibly inappropriate for the process at hand, and attempt to quantify the
uncertainty
associated with the choice of an appropriate value.
Given the SNO  experimental error, this theoretical uncertainty is unlikely
to affect
the interpretation of the CC results. It may, however, be relevant to
future, more
precise determinations of Gamow-Teller transitions in other contexts.

Finally, for completeness, we revisit the analysis of the NC radiative
correction
computed in Ref. \cite{Towner}. In this case, bremsstrahlung contributions are
highly suppressed, the correction is governed by virtual gauge boson exchange,
and the result is esssentially $E_\nu$-independent. We obtain a correction
to the NC
cross section that is a factor of four larger than given in Ref.
\cite{Towner}, which
neglected the dominant graph. The implication of a complete analysis is the
application of a $\sim 1.5\%$ correction to the tree-level NC cross section.

Our discussion of these points is organized in the remainder of the paper
as follows.
In Section II we present our formalism for the CC radiative corrections.
Given the thorough discussion of this formalism in Ref. \cite{Towner}, we
restrict ourselves
to only brief explanation of the basic formalism that is used to evaluate
the corresponding Feynman graphs and deduce the formulae for
the differential cross section.
In Section III we discuss the delicate issue of bremsstrahlung thresholds 
and the `detector dependence' of the  CC
radiative corrections. In particular,
we derive the corrections for two extreme cases of very high and very low
thresholds, and for an intermediate, more realistic case \cite{Hamish}.
We show in Section III where our results disagree with those of Ref.
\cite{Towner}, and trace the origin of these discrepancies.
Detailed tabular evaluation of the modification
of the differential CC cross section for the `realistic' bremsstrahlung
threshold is provided as well.
In Section IV we consider the effects of the electron spectrum
distortion. In particular, we consider the test of
the oscillation null hypothesis, where the unperturbed $\nu_e$ spectrum
of the $^8$B decay is expected.
In Section V we derive the corrections to the NC
reaction rate and discuss the differences with their
treatment in \cite{Towner}. We conclude in Section VI. Finally,
in Appendix  we collect needed formulae for the evaluation
of the triple differential cross section (in $E_e, E_{\gamma}$ and the angle
between them) for an arbitrary  bremsstrahlung threshold.


\section{General considerations}

In the charged current neutrino disintegration of deuterons at rest
in the laboratory frame,
Eq.(\ref{eq:CC}),
the incoming $\nu_e$ energy $E_{\nu}$,
corrected for the mass difference
$\Delta = M_d - 2M_p = -0.931$ MeV, is shared by the outgoing electron
(energy $E_e$), the energy of the relative motion of the two protons
$p^2/M_p$, and by the energy of a bremsstrahlung photon $E_{\gamma}$
(if such a photon is emitted), i.e.,
\begin{equation}
E_{\nu} + \Delta = E_e + p^2/M_p + (E_{\gamma})~.
\label{eq:econs}
\end{equation}
This energy conservation condition must be, naturally, always obeyed.
For the neutrino energies we are
considering the motion of the center of mass of the protons
can be neglected.

Since radiative corrections are only few percent in magnitude, we
follow Towner \cite{Towner} and use for the `tree level'
differential cross section the formula based on the effective
range theory (see \cite{Kelly,Ellis})
\begin{equation}
\label{eq:TreeCrossSect}
\left( \frac{{\rm d}\sigma_{CC}}{{\rm d}E_e} \right)_{tree} =
\frac{2G_F^2}{\pi} V_{ud}^2 g_A^2 M_p p_e E_e p |I(p^2)|^2 ,
\end{equation}
where for $p^2$ we should substitute $M_p(E_{\nu} + \Delta - E_e)$.
It is important to remember that the radial integral,
the overlap of the radial wave function of the two continuum
protons and the bound state
\begin{equation}
\label{eq:Integral}
I(p^2) = \int u_{cont}^*(pr)u_d(r)dr
\end{equation}
also depends on the momentum $p$ of the relative motion of the two
protons.

\begin{figure}
\begin{center}
\includegraphics[width=5.5in]{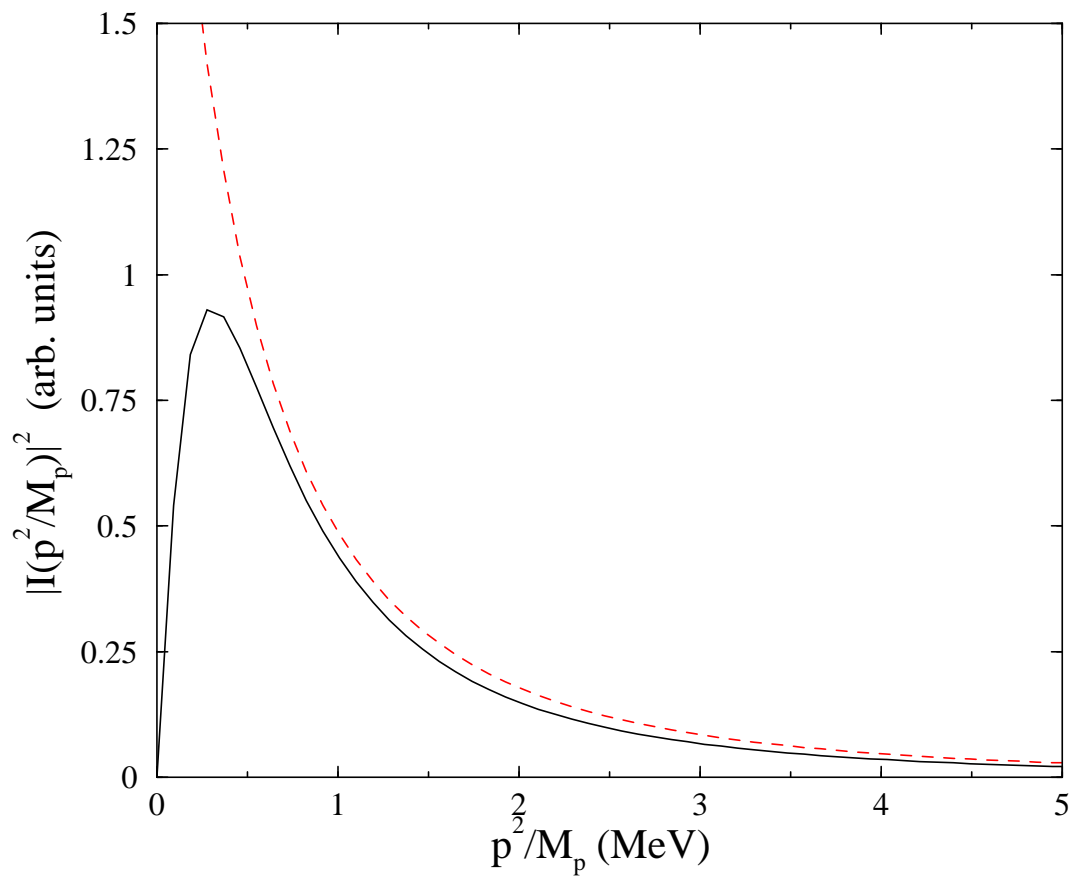}
\end{center}
\caption{Radial integral $|I(p^2/M_p)|^2$.
Exact result (solid line) and scaled approximation with Coulomb repulsion and
effective range set to zero (dashed line) are shown. }
\label{fig:isquare}
\end{figure}

We plot in Fig. \ref{fig:isquare} the quantity $|I(p^2)|^2$ evaluated
as in Ref. \cite{Kelly}, i.e., using the scattering length and effective
range approximation as well as the Coulomb repulsion of the
two final protons. The most important feature of the $p^2$ dependence
is its width  when expressed in the relevant units
of $p^2/M_p$, the kinetic energy of the continuum protons. It is easy
to understand the width of the curve as demonstrated in the figure.
The dashed line represents the same $|I(p^2)|^2$ evaluated neglecting  Coulomb
repulsion as well as effective range. In that case a simple analytic
expression obtains,
\begin{equation}
\label{eq:IntAnalit}
|I(p^2)|^2 \simeq \frac{const}{(1 + a_{pp}^2 p^2)(1 + \frac{p^2}{E_b M_p})^2} ~,
\end{equation}
where $E_b$ is the deuteron binding energy and the proton-proton 
scattering length is $a_{pp}$ = -7.82 fm.
The value of the proportionality constant is irrelevant 
in the present context.
Thus the width is  determined essentially by
$ \sim (\hbar c)^2/(a_{pp}^2 M_p) \sim$ 0.7 MeV (the term with 
$p^2/E_b M_p$ contributes very little to the width)
in agreement with the more accurate evaluation. We explain the
relevance of this width later, in Section III.

\begin{figure}
\hspace{0.00in}
\begin{center}
\includegraphics[width=5in]{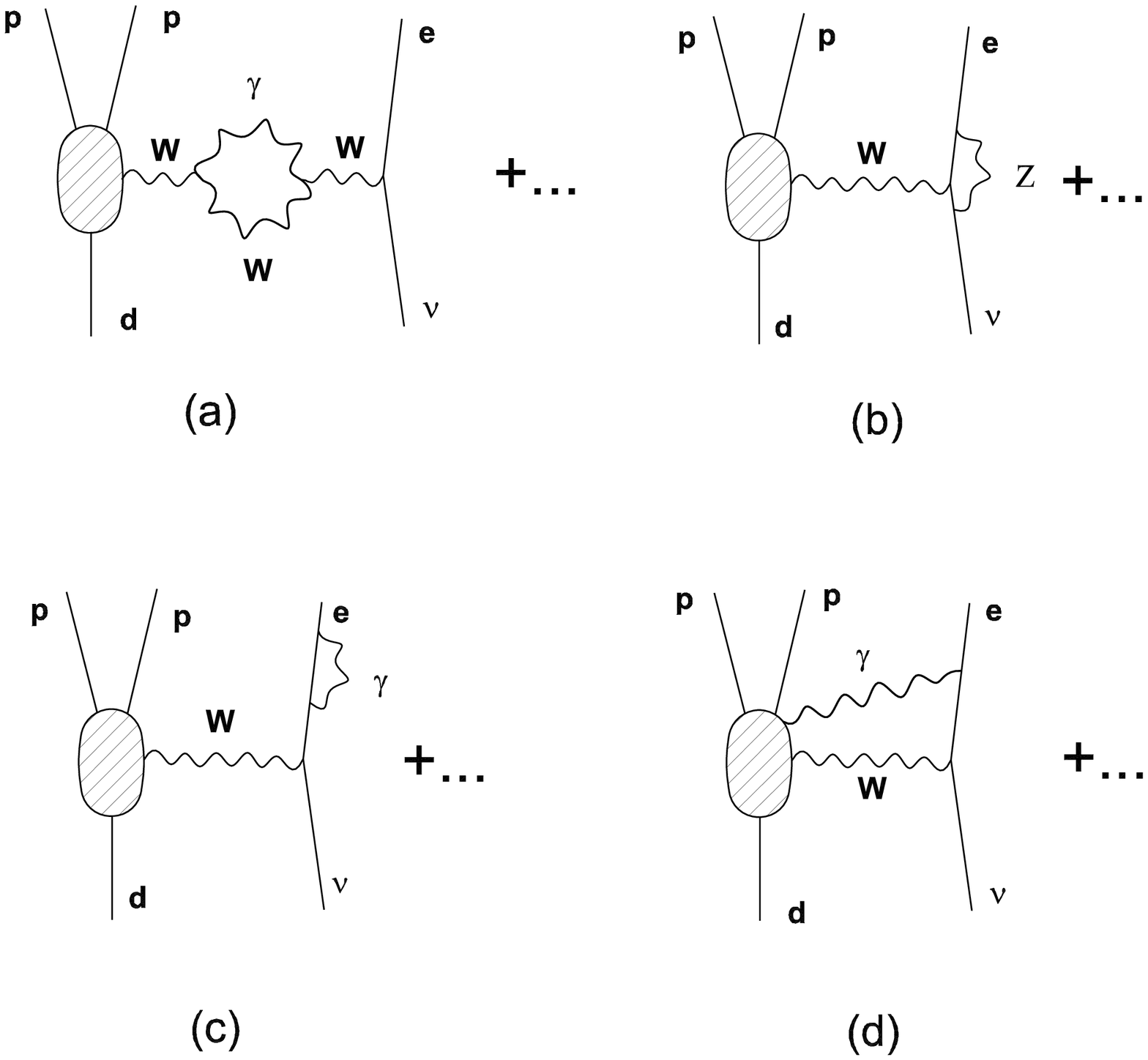}
\end{center}
\caption {Order $\alpha$ radiative corrections to the
charged-current breakup of deuteron: $d+\nu_e \rightarrow p+p+e$
involving virtual $\gamma$ quanta and $Z$ boson exchange. 
The large shaded oval represents the vertex with all its hadronic
complications.}
\label{fig:graph1}
\end{figure}

\begin{figure}
\hspace{0.00in}
\begin{center}
\includegraphics[width=4.5in]{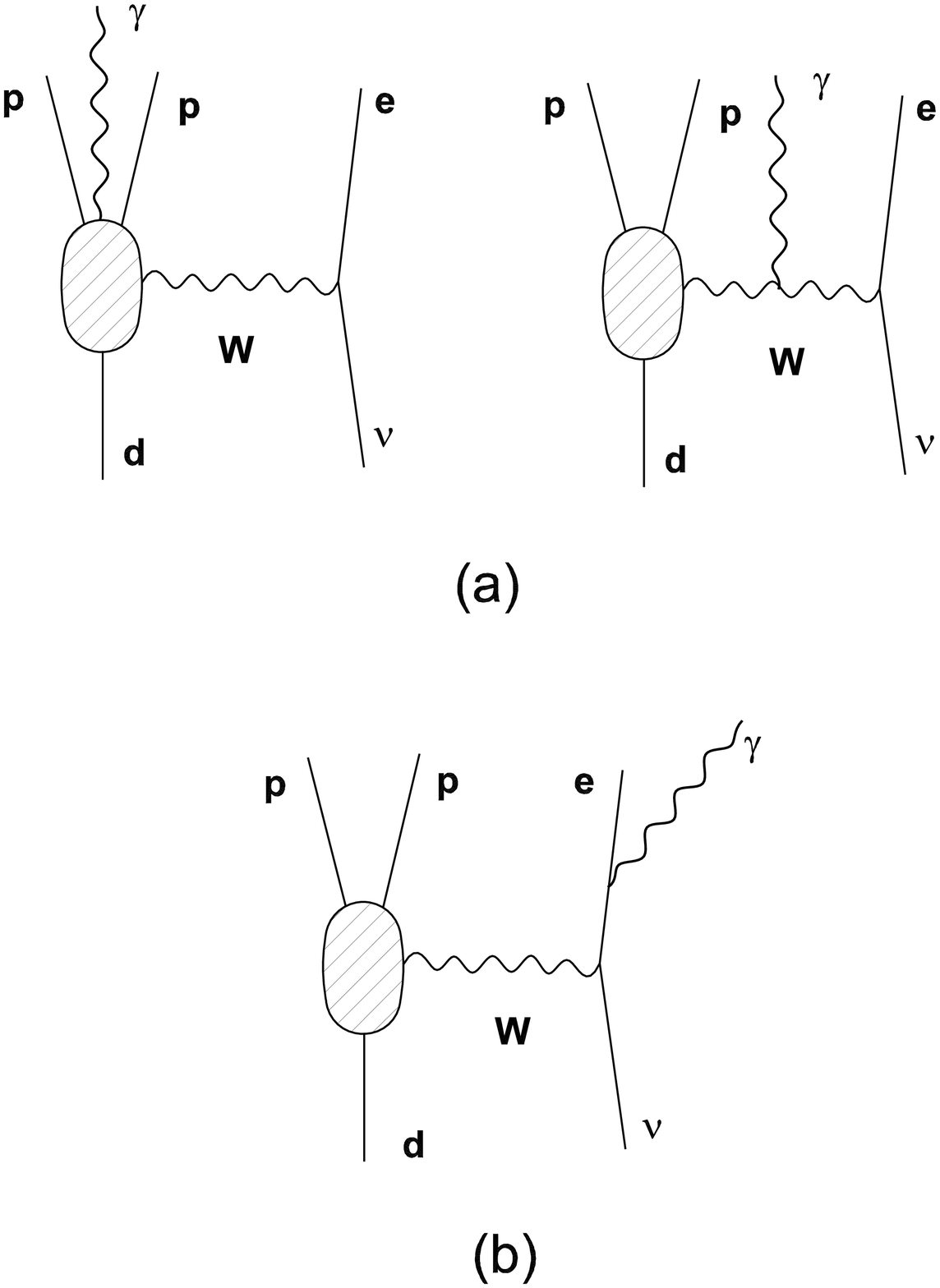}
\end{center}
\caption{Order $\alpha$  corrections due to bremsstrahlung emission.
See caption to Fig. \protect\ref{fig:graph1}. }
\label{fig:graph2}
\end{figure}

The radiative corrections consist of two components: the exchange
of virtual photons and $Z$-bosons
and the emission of real bremsstrahlung photons.
The Feynman graphs for the exchange of virtual $\gamma$ quanta
and $Z$ bosons are shown in Fig. \ref{fig:graph1}.
The bremsstrahlung graps are shown in Fig. \ref{fig:graph2}.
The photon emission by the moving electron is dominant ( graph (b) in
Fig. \ref{fig:graph2}),
but the complete set of graphs must be considered for
gauge invariance.
The treatment of radiative
corrections proceeds along the well tested lines developed
for the treatment of beta decay (see Ref. \cite{Sirlin} for a review).

Let us consider the virtual exchange corrections first. While the treatment of
corrections involving only leptons is straightforward, those involving
hadronic participants
require considerable care. To that end, it is useful to adopt the framework
of an EFT, valid
below a scale $\mu\sim 1$ GeV. Long-distance physics ($p\lesssim\mu$)
associated with
non-perturbative strong interactions are subsumed into hadronic matrix
elements of appropriate
hadronic operators. Short distance physics ($p\gtrsim\mu$) contributions
are contained in
coefficient functions $C(\mu)$ multiplying the effective operators (see,
{\em e.g.}, the
discussion in Ref. \cite{Ji:1991}). In the present case, the CC reaction of Eq.
(\ref{eq:ccreaction}) is dominated by the pure Gamow-Teller transition
$^3$S$_1\to^1$S$_0$.
Thus, for the low-energy EFT, we require matrix elements of the effective,
hadronic axial
current. The resulting CC amplitude is
\begin{equation}
\label{eq:ccampl}
M(^3{S}_1\to^1{S}_0) = -{G_F\over\sqrt{2}}V_{ud} {\bar
e}\gamma_\lambda
(1-\gamma_5) \nu C(\mu) \langle ^1{S}_0| {\tilde A}^\lambda
|^3{S}_1\rangle+\cdots\ \ ,
\end{equation}
where $C(\mu)$ is the short-distance coefficient function mentioned above;
${\tilde A}^\lambda$
is an effective, isovector axial current operator built out of low-energy
degrees of freedom
({\em e.g.}, nucleon and pion fields); the $+\cdots$ denote contributions
from higher-order
effective operators;  and where the $\mu$-dependence of $C(\mu)$
compensates for that of the axial current matrix element, leading to a
$\mu$-independent
result.  In effect, the presence of $C(\mu)$ is needed  for matching of
the effective theory
onto the full theory (QCD plus the electroweak Standard Model).

Note that we have normalized the amplitude to the Fermi constant determined
from
the muon lifetime\footnote{This value is sometimes denoted $G_\mu$ in the
literature.}, $G_F = 1.16639(5)\times 10^{-5}$ GeV$^{-2}$ \cite{PDG}. Thus,
$C(\mu)$ contains the difference
\begin{equation}
\label{eq:diff}
\Delta r_\beta^{A(p \ge \mu)} - \Delta r_\mu\ \ ,
\end{equation}
where $\Delta r_\beta^{A(p \ge \mu)}$ contains the 
short distance virtual corrections to the axial vector
semileptonic amplitude and $\Delta r_\mu$ denotes the Standard Model
electroweak
radiative corrections to the muon decay amplitude. In the difference
(\ref{eq:diff}),
all universal short-distance effects (Fig. 2a-c) cancel, leaving only
contributions
from the non-universal parts of  diagrams in Fig. 2b-d.

As a corollary, we emphasize that care must be exercised in choosing a
value for the
axial coupling constant 
$g_A$ used in computing $ \langle ^1{S}_0| {\tilde A}^\lambda
|^3{S}_1\rangle$. Typically, $g_A$ is determined from the experimental ratio
\cite{hardy}
\begin{equation}
\label{eq:gadef}
\lambda = {G_A'\over G_V'} = {G_A(1+\Delta r_\beta^A)\over G_V(1+\Delta
r_\beta^V)}
\approx {G_A\over G_V}(1+\Delta r_\beta^A-\Delta r_\beta^V)  \ ,
\end{equation}
where $\Delta r_\beta^V$ ($\Delta r_\beta^A$) denotes the 
total radiative correction to the vector (axial vector)
semileptonic
amplitude. The CVC relation implies $G_V=G_F V_{ud}$, while the axial 
 coupling constant is
{\em defined} via $G_A=g_A G_F V_{ud}$. To the extent that $\Delta r_\beta^V=
\Delta r_\beta^A$, the ratio $\lambda$ is just $g_A$. As we note below,
however,
hadronic contributions to $\Delta r_\beta^V$ and $\Delta r_\beta^A$ are 
in general not identical.
While we speculate that the differences are considerably smaller than
relevant here,
arriving at a reasonable  estimate requires a future, more systematic study.

The asymptotic (short-distance) contributions to $C(\mu)$ have been
computed in Ref.
\cite{Sirlin} using current algebra techniques and the short-distance
operator product
expansion. The result implies
\begin{equation}
\label{eq:shortdist}
C(\mu) = 1 + {\alpha\over 2\pi}\left[3{\bar
Q}\ln\frac{M_Z}{\mu}+\frac{3}{2}\ln\frac{M_Z}{\mu}+
\frac{1}{2}{\cal A}_g(\mu) \right] + b(\mu)\ \ ,
\end{equation}
where ${\bar Q}$ is the average charge of the quarks involved in the transition
\begin{equation}
{\bar Q} =\frac{1}{2}(Q_u+Q_d)=\frac{1}{6} \ \ ,
\end{equation}
${\cal A}_g(\mu)$ contains short-distance QCD corrections,
and $b(\mu)$ must be included to correct for any mismatch between the
$\mu$-dependence
appearing elsewhere in $C(\mu)$ and that appearing in the matrix element of
${\tilde
A}_\lambda$. Explicit expressions for the short-distance QCD contributions
${\cal
A}_g(\mu)$ may be found in Ref. \cite{Sirlin}. We note that the second term
of Eq.
(\ref{eq:shortdist}) ($\propto{\bar Q}$) arises from the sum of box
diagrams involving
$(\gamma, W)$ and $(Z,W)$ pairs, while the third term arises from QED
external leg and vertex
corrections. When long-distance, ${\cal O}(\alpha)$ virtual effects arising
from the matrix
elements in Eq. (\ref{eq:ccampl}) are included along with those appearing
in $C(\mu)$, the
$\mu$-dependence of the third term in Eq. (\ref{eq:shortdist}) cancels
completely.

Long-distance virtual photon contributions also contain an infrared
singularity which is
conventionally regulated by including a photon \lq\lq mass" $\lambda$. The
resulting
$\lambda$-dependence is cancelled by corresponding $\lambda$-dependence in
the bremsstrahlung
cross section, yielding a $\lambda$-independent correction to the total CC
cross section.
In what follows, then, it is convenient to consider the ${\cal O}(\alpha)$
correction to
the tree-level cross section:
\begin{equation}
\label{eq:sigcor}
d\sigma_{CC} = d\sigma_{CC}^{tree}\left[1+\frac{\alpha}{\pi}g\right] ~,
\end{equation}
where the correction factor $g$ depends on $E_\nu$ and $E_e$ as well as on
$E_\gamma$
when bremsstrahlung photons are detected. This function receives
contributions from $C(\mu)$,
\begin{equation}
\label{eq:gvshort}
\frac{\alpha}{\pi} g_v^{p\gtrsim\mu}=2\left[C(\mu)-1\right] \ ,
\end{equation}
long-distance ($p\lesssim\mu$) virtual contributions to the
the axial current matrix element in Eq. (\ref{eq:ccampl}),
$g_v^{p\lesssim\mu}$, and the
bremsstrahlung differential cross section, $g_b$.

In the analysis of Ref. \cite{Towner}, the long-distance
contributions arising from virtual processes is obtained by treating the
nucleon
as a point-like, relativistic particle. The result is
\begin{eqnarray}
g_v^{p\lesssim\mu}& = & \frac{3}{2} {\rm ln} \left( \frac{\mu}{M_p} \right)
+ 3{\bar Q} {\rm ln} \left( \frac{\mu}{M_A} \right)+
 {\cal A}-\frac{3}{8}
 \nonumber \\
{\cal A} & = & \frac{1}{2} \beta {\rm ln} \left( \frac{1 + \beta}{1 - \beta}
\right) -1 + 2 {\rm ln}  \left( \frac{ \lambda } { m_e } \right)
\left[ \frac{1}{2 \beta }
{\rm ln} \left( \frac{1 + \beta}{1 - \beta} \right)
 - 1 \right] \nonumber \\
& & + \frac{3}{2} {\rm ln}  \left( \frac{ M_p } { m_e } \right)
- \frac{1}{ \beta }
\left[ \frac{1}{2} {\rm ln} \left( \frac{1 + \beta}{1 - \beta} \right)
\right]^2
+ \frac{1}{ \beta } L \left( \frac{ 2 \beta }{ 1 + \beta } \right) ~.
\end{eqnarray}
Here, $L$ is  the Spence function and the $-3/8$ is added
to obtain agreement with the $\beta$-decay correction \cite{Sirlin} and
neutrino
capture reaction $\bar{\nu_e} + p \rightarrow e^+ + n$ as calculated in Refs.
\cite{vogel,fayans}. Note that when this $-3/8$ is added to ${\cal A}$, the
resulting
expression agrees with the calculations of Refs. \cite{Sirlin,vogel,fayans}.

We observe
that the sum $g_v^{p\gtrsim\mu}+g_v^{p\lesssim\mu}$ is independent of
$M_p$. It does,
however, contain the logarithm
\begin{equation}
\label{eq:malog}
3{\bar Q}\ln\frac{M_Z}{M_A}
\end{equation}
where $M_A$ has been chosen in Ref. \cite{Towner} as a hadronic scale
associated with the
long-distance part of the $(W,\gamma)$ box diagram. Neglecting terms
proportional to $E_e$ and
$m_e$, the sum of the box and crossed-box diagrams depends on the
antisymmetric T-product of
currents
\begin{equation}
\label{eq:boxes}
\epsilon_{\mu\nu\lambda\rho}\int\ d^4xe^{ik\cdot x} \langle ^1{S}_0|
T[J^\lambda_{EM}(x) J^\rho_{CC}(0)] |^3{S}_1\rangle \ \ \ ,
\end{equation}
where $J^\lambda_{EM}$ and $J^\rho_{CC}$ denote the electromagnetic and
weak charged
currents, respectively, and where the $\mu$ and $\nu$ indices are
contracted with loop
momentum and the lepton current. In order that the antisymmetric T-product
appearing in
Eq. (\ref{eq:boxes}) produce a Gamow-Teller transition, only the vector
current part of
$J^\rho_{CC}$ must be retained. In contrast, for pure Fermi transitions as
considered in
Ref. \cite{Sirlin}, only the axial vector charged current operator
contributes. In that work,
a choice for the hadronic scale was made based by considering $\pi$
$\beta$-decay (a pure
Fermi transition), and a vector meson dominance model for the axial vector
charged current
operator, leading to the appearance of the $a_1$
meson mass $M_A$ as the long-distance hadronic scale.

In the present case, such a choice appears inappropriate, since the relevant
current operator is
a vector, rather than axial vector current. To the extent that the vector
meson dominance
picture is as applicable to nucleons as to pions, a more reasonable choice
for the hadronic
scale would be $m_\rho$. However, such a choice is unabashedly
model-dependent and calls for
some estimate of the theoretical uncertainty. On general grounds, it is
certainly reasonable
to choose a hadronic scale anywhere between the chiral scale
$\Lambda_\chi=4\pi F_\pi\approx$
1.17 GeV and $\Lambda_{QCD}\approx$ 200 MeV. Indeed, the latter
choice could arise
naturally from $\Delta$-intermediate state contributions to the
$(W,\gamma)$ box diagrams.
Thus, we replace the logarithm in Eq. (\ref{eq:malog})
\begin{equation}
3{\bar Q}\ln\frac{M_Z}{M_{had}}= 2.39_{-.21}^{+.67}  \ ,
\end{equation}
where the central value corresponds to $M_{had}=m_\rho$; the upper
value corresponds
to  $M_{had}=\Lambda_{QCD}$; and the lower value is obtained with
$M_{had}=\Lambda_\chi$. This range corresponds to a spread of 0.2\%
in predictions for the cross section
\footnote{We note that a similar estimate of the
hadronic uncertainty in the box contributions to the Fermi amplitude was
made in Ref. \cite{hardy}.}. While this uncertainty is too small to affect the
determination of the $^8$B neutrino flux, it could affect more precise
determinations of Gamow-Teller transitions for other purposes.

The choice of $M_{had}$ amounts to use of a model for $b(\mu)$. A
source of potentially
larger theoretical uncertainties lies in possible additional,
model-dependent contributions to
this constant. While a complete study of these effects goes beyond the
scope of the present
work, we observe that the hadronic uncertainty cannot be finessed away
using, {\em e.g.},
chiral perturbation theory, since we have no independent measurements from
which to fix the
relevant low-energy constants. Moreover, the $\mu$-dependence introduced
through the
short-distance QCD correction ${\cal A}_g(\mu)$ must be cancelled by a
corresponding
$\mu$-dependent term in $b(\mu)$. To date, no calculation has produced such
a cancellation.
While the effect of this uncorrected mismatch between short- and
long-stance effects is
likely to be small, we are unable to quantify it at the present time.

In contrast to the virtual corrections, the bremsstrahlung correction $g_b$ is
relatively free from hadronic uncertainties.
In order to evaluate the bremsstrahlung part, one has to add, in principle,
the contribution of all graphs with photon lines attached to all
external charged particles. Only the sum of these graphs is
gauge-invariant. However, for the low energies, the
electron bremsstrahlung dominates over the proton, deuteron,
and $W$ bremsstrahlung.

Writing again the correction to the cross
section in the form $1 + \alpha/\pi g_b(E_e, E_{\nu})$
one obtains the differential bremsstrahlung correction in the form
\begin{eqnarray}
\label{eq:Bremst-gb-function}
\frac{{\rm d} g_b(E_e, E_{\nu},k)}{{\rm d} k} & = &
\left[\frac{ E_{\nu} + \Delta - E_e - E_{\gamma}}{ E_{\nu} + \Delta - E_e}
\right]^{1/2} \frac{k^2}{2 E_{\gamma}} \nonumber \\
& \times  & \int_{-1}^{+1} {\rm d} x \left[ \frac{E_{\gamma}}
{E_e^2 (E_{\gamma} - \beta k x)} +
\beta^2 \frac{E_e + E_{\gamma}}{E_e} \frac{1 - k^2 x^2/E_{\gamma}^2}
{(E_{\gamma} - \beta k x)^2} \right] ~,
\end{eqnarray}
where $k$ is the photon momentum, and
$E_{\gamma} = (k^2 + \lambda^2)^{1/2}$, i.e. $\lambda$ is as before
the `photon mass'. Also, $x = \cos( \theta_{e,\gamma} )$.

The dependence on the `photon mass' $\lambda$ is eliminated only
when one adds to the  $\lambda$ dependent part of the virtual
correction $\cal A$ an
integral over the bremsstrahlung spectrum up to some
$E_{\gamma}^{min} \gg \lambda$. We will discuss the various possible
choices of $E_{\gamma}^{min}$ in the next section, but here as an
example we evaluate one of the integrals that appears in that context
\begin{eqnarray}
\label{eq:example}
& & \int_0^{E_{\gamma}^{min}} \frac{k^2 {\rm d} k}{\sqrt{\lambda^2 + k^2}}
\int_{-1}^{+1}  \frac{{\rm d} x}{(E_{\gamma} - \beta k x)^2} =
2 \int_0^{E_{\gamma}^{min}} \frac{k^2 {\rm d} k}{\sqrt{\lambda^2 + k^2}
(\lambda^2 + m_e^2/E_e^2 k^2) }
\nonumber \\
& = & 2 \frac{E_e^2}{m_e^2} \left[  \int_0^{E_{\gamma}^{min}}
\frac{ {\rm d} k}{\sqrt{\lambda^2 + k^2}} -
 \int_0^{E_{\gamma}^{min}} \frac{\lambda^2  {\rm d} k}{\sqrt{\lambda^2 + k^2}
(\lambda^2 + m_e^2/E_e^2 k^2)} \right]
\nonumber \\
& = &  2 \frac{E_e^2}{m_e^2} \left[ {\rm ln}\frac{2 E_{\gamma}^{min}}{\lambda}
- \frac{1}{2 \beta}  {\rm ln}  \left( \frac{1 + \beta}{1 - \beta} \right)
\right] ~,
\end{eqnarray}
where $\beta = p_e/E_e$ and  the last integral,
which is independent of $E_{\gamma}^{min}$,  was evaluated
after the substition $z = k/\lambda$ in the
limit $E_{\gamma}^{min}/\lambda \rightarrow \infty$. One must not
use the  limit  $E_{\gamma}^{min} \rightarrow 0$ before
all terms containing $\lambda$ are eliminated.

To evaluate the full radiative correction, we assume that
in an experiment one measures the number of events with energy $E_{obs} \pm
{\rm d}E_{obs}$.
Here $E_{obs} = E_e$ when the bremsstrahlung photon (if such a photon
is emitted) has energy of less than $E_{\gamma}^{min}$. We will also
assume that when $E_{\gamma} \ge E_{\gamma}^{min}$ then $E_{obs} = E_e +
E_{\gamma}$.
(In the next section we will also consider a modification to the latter rule,
making it closer to the actual conditions of the SNO
experiment \cite{Hamish}.)

Thus the radiative correction to the cross section can be expressed as
\begin{equation}
\label{eq:TotalRadCorr}
\left( \frac{{\rm d}\sigma_{CC}}{{\rm d}E_{obs}} \right)_{rad} =
\frac{\alpha}{\pi}
\left[
g_v  + g_b^{low}(E_{\gamma} <  E_{\gamma}^{min}) + 
g_b^{high}(E_{\gamma} \ge E_{\gamma}^{min})
\right] ~.
\end{equation}
We describe in the next section how to evaluate these three functions in
general
and for three particular choices of $ E_{\gamma}^{min}$.

\section{Radiative corrections to the CC cross section}

Treatment of radiative corrections involving virtual photon exchange as well as
bremsstrahlung  photon
emission is a delicate issue due to the appearence of infrared divergencies.
In our analysis we follow the conventional approach of introducing
infrared regulator in the form of photon mass $\lambda$, and split
the bremsstrahlung contributions into two pieces below
and above the threshold value $E_{\gamma}^{min}$ as explained above.
When contribution from virtual photon exchange are added to the piece
with $E_{\gamma}<E_{\gamma}^{min}$,
the dependence on infrared regulator $\lambda$ is eliminated.
However, it is effectively replaced by a dependence on $E_{\gamma}^{min}$.

The threshold $E_{\gamma}^{min}$ is a detector-dependent quantity, and
may vary depending on the experimental conditions.
In addition, the experimental conditions also dictate how to combine the piece
with $E_{\gamma} < E_{\gamma}^{min}$ ( $g_v + g_b^{low}$)
and the $E_{\gamma} > E_{\gamma}^{min}$ part. Thus it is impossible to
give a completely general recipe here.

With this caveat in mind,
in our analysis we adopt a following framework. Each
detected CC event is characterized by the recorded energy $E_{obs}$
which in general is a function of electron energy $E_{e}$ and,
if present, photon energy $E_{\gamma}$: $E_{obs}={E_{obs}}(E_{e},E_{\gamma})$.
In the following we concentrate in particular on the role played in this context by
the threshold $E_{\gamma}^{min}$.

We consider the following situations:

A) The electrons are always recorded above the electron detection threshold $E_{e}^{min}$,
and the bremsstrahlung photons are never detected, i.e. $E_{\gamma}^{min} \rightarrow \infty$.

B) The electrons are always recorded above the electron detection threshold $E_{e}^{min}$,
and the bremsstrahlung photons are also always detected, i.e.
 $E_{\gamma}^{min} \rightarrow 0$.

C) A more realistic case, resembling the actual situation in the SNO
detector \cite{Hamish} when only part of the $E_{\gamma}$ energy is recorded, namely
$E_{obs}=(E_{e} - m_e)\theta(E_{e}  - E_{e}^{min}) + m_e +(E_{\gamma}-E_{\gamma}^{min})
\theta(E_{\gamma}-E_{\gamma}^{min})$. Here $\theta (x)$ is the step function.

We simplify the cases A) and B) even further by considering an idealized detector
with $E_{e}^{min} = m_e$, i.e. all electrons and, thus, all neutrino interaction
events are detected. When integrated over $E_{obs}$ one arrives at the total number
of events caused  by a neutrino of energy $E_{\nu}$. That quantity
must be, naturally, independent on the bremsstrahlung threshold $E_{\gamma}^{min}$.
This is the consistency requirement imposed by Beacom and Parke \cite{John}.
We verify that our results fulfill this condition.

In Fig. \ref{fig:differential} we plot as an example the normalized 
radiative correction to the differential
cross-section ${{\delta d\sigma(E_{\nu},E_{obs})} \over {dE_{obs}}} / {\sigma_{tot}^{tree}(E_\nu)}$
for $E_{\nu}$ = 10 MeV and two extreme cases $E_{\gamma}^{min}\rightarrow \infty$
(bremsstrahlung never detected, full line) 
and $E_{\gamma}^{min}\rightarrow 0$ (bremsstrahlung always detected, dashed line).
The two corresponding curves are quite different, reflecting the different
dependence of $E_{obs}$ on $E_e$ and $E_{\gamma}$. However, the areas under the curves
are equal as they must for consistency.

It is interesting to note that evaluation by Towner \cite{Towner}
considers the same limiting cases. However,
the results of Ref. \cite{Towner} give different corrections to
the total cross section, $\delta\sigma^{tot}$,
i.e. they fail  the consistency check. In fact, our results and Ref. \cite{Towner}
differ in both extremes.
We trace now the origin of these discrepancies.

\subsection{The case of  $E_{\gamma}^{min}\rightarrow \infty$, no bremsstrahlung
detected}

Let us first consider the limit $E_{\gamma}^{min}\rightarrow \infty$. 
In that case we have to integrate
the bremsstrahlung spectrum over the photon momentum from $0 \rightarrow \infty$.
At the same time, the energy conservation condition, Eq. (\ref{eq:econs}), 
must be obeyed. Since now $E_{obs} \equiv E_e$, then
for a fixed $E_{\nu} + \Delta - E_e$ the quantity $p^2/M_p$ must be varied
together with $E_{\gamma}$. As noted above, 
and illustrated in Fig. \ref{fig:isquare}, the quantity $|I(p^2/M_p)|^2$ is a rapidly
varying function which falls off quickly for $p^2/M_p \ge$ 0.7 MeV. Therefore to
correctly account for this dependence we write

\begin{equation}
\label{eq:BremstCorrect}
\left( \frac{{\rm d}\sigma_{CC}}{{\rm d}E_e}
\right)_{b} ={\alpha \over \pi} \left( \frac{{\rm d}\sigma_{CC}}{{\rm d}E_e}
\right)_{tree}  \int_0^{E_{\gamma}^{min}} {{|I(E_{\nu}+\Delta-E_e-E_{\gamma}^{k})|^2}\over 
{|I(E_{\nu}+\Delta-E_e)|^2}} 
\frac{{\rm d} g_b(E_e, E_{\nu}, k)}{{\rm d}k}dk,
\end{equation}
where $g_b(E_e, E_{\nu},k)$ is given in Eq. (\ref{eq:Bremst-gb-function}).
If $|I(E_{\nu}+\Delta-E_e-E_{\gamma})|^2$ could in fact be treated as a constant,
the ratio of the two  $I^2$ would be unity, and Eq. (\ref{eq:BremstCorrect})
would be identical to Eq. (13) in \cite{Towner}.
To make the connection with Ref. \cite{Towner} even more concrete we write
(note that for $E_{\gamma}<E_{\gamma}^{min}$, $E_{obs}=E_e$):

\begin{eqnarray}
\label{eq:BremstCorrect-Transformed}
\left( \frac{{\rm d}\sigma_{CC}}{{\rm d}E_{obs}} \right)_{b} & = &
\left( \frac{{\rm d}\sigma_{CC}}{{\rm d}E_{obs}} \right)_{b}^{[Towner]} +  
{\alpha \over \pi} \left( \frac{{\rm d}\sigma_{CC}}{{\rm d}E_{obs}} \right)_{tree}  
\nonumber \\
& \times & \int_0^{E_{\gamma}^{min}} \left({{|I(E_{\nu} + \Delta - E_{obs} - E_{\gamma}^{k})|^2}
\over {|I(E_{\nu} + \Delta - E_{obs})|^2}} - 1\right) 
\frac{{\rm d} g_b(E_{obs}, E_{\nu}, k)}{{\rm d} k}dk ~.
\end{eqnarray}
(In Eqs. (\ref{eq:BremstCorrect},\ref{eq:BremstCorrect-Transformed}) the upper limit
of the integral obviously should not extend beyond the corresponding bremssstrahlung
endpoint.)

The first term on the r.h.s of Eq. (\ref{eq:BremstCorrect-Transformed}) 
is the contribution present in Ref. \cite{Towner}. 
It contains infrared divergence that  disappears after contributions 
from virtual photons are added. The second term is infrared finite. 
Due to the shape of $|I(E_{\nu} + \Delta - E_e - E_{\gamma}^k)|^2$ 
this term enhances the contribution of the low energy tail in 
$ ({\rm d}\sigma_{CC}/{\rm d}E_{obs})_{b}$. 
The overall result of the low $E_{obs}$ tail enhancement is that the 
total cross-section is increased by about 3\% compared to corresponding result in \cite{Towner}
for the considered case of $E_{\nu}$ = 10 MeV.

\subsection{ The case  $E_\gamma^{min} \rightarrow 0$, bremsstrahlung always
detected}

If one wants
to study the opposite extreme $E_\gamma^{min} \rightarrow 0$ it is crucial in
Eq. (\ref{eq:TotalRadCorr}) to first add all three terms, eliminate infrared
cutoff dependence, and only then take the limit  $E_{\gamma}^{min} \rightarrow 0$. 
The order of limits $\lambda \rightarrow 0$ and $E_{\gamma}^{min} \rightarrow
0$ is important because upper limit of integrals like Eq. (\ref{eq:example}) is
$E_{\gamma}^{min}/{\lambda}$. Since $\lambda$ ultimately is an
infinitesimal unphysical parameter it is mandatory to maintain
$E_{\gamma}^{min} \gg \lambda$ during the entire course of the calculation
\footnote{If $\lambda$ were truly the photon mass, the
requirement that $E_{\gamma} > \lambda$ would be obvious.}.
This leads to a non-intuitive result that the second term in Eq.
(\ref{eq:TotalRadCorr}) has a non-zero contribution even in the limit
$E_{\gamma}^{min} \rightarrow 0$. In particular, one must write the
following expression corresponding to the second term on the r.h.s of Eq.
(\ref{eq:TotalRadCorr}), i.e. for  $E_{\gamma} \le E_{\gamma}^{min}$:

\begin{equation}
\label{eq:Bremst-Zero-Threshhold}
\left( \frac{{\rm d}\sigma_{CC}}{{\rm d}E_{obs}} \right)_{b} =\left(
\frac{{\rm d}\sigma_{CC}}{{\rm d}E_{obs}} \right)_{tree}{{\alpha} \over
{\pi}} \left( 2 \ln \left( {{E_{\gamma}^{min}}\over{\lambda}} \right)
\left[ \frac{1}{2 \beta}  \ln  \left( \frac{1 + \beta}{1 - \beta} \right)
-1 \right] + {\cal C}(\beta) \right) + {\cal
O}(E_{\gamma}^{min})
\end{equation}

with

\begin{eqnarray}
\label{eq:CFunction}
& & {\cal C}(\beta)=2\ln(2)\left[
\frac{1}{2 \beta}  \ln  \left( \frac{1 + \beta}{1 - \beta} \right) -1
\right]+1+\frac{1}{4 \beta}  \ln  \left( \frac{1 + \beta}{1 - \beta}
\right) \nonumber \\
& & \times \left[ 2+ \ln \left( {{1-\beta^2} \over 4}
\right) \right] + {1 \over {\beta}} \left[ {L(\beta)-L(-\beta)} \right] +
{1 \over 2 \beta} \left( {L\left( {{1-\beta} \over {2}} \right) - L\left(
{{1+\beta} \over {2}} \right)} \right)~, 
\end{eqnarray}
where for $\beta < 1$ the Spence function is
\begin{equation}
L(\beta)=\int_0^{\beta}{{\ln(| 1-x |)}\over x}dx = 
- \sum_{k=1}^{\infty}\frac{\beta^k}{k^2} ~.
\end{equation}

The $\lambda$-dependent terms in Eq.
(\ref{eq:Bremst-Zero-Threshhold}) will be cancelled by $\lambda$-dependent
pieces from virtual photon contributions, and the logarithmic divergence in
$E_{\gamma}^{min}$ will disappear after the piece with
$E_{\gamma}>E_{\gamma}^{min}$ is added to the cross-section (third term in
Eq. (\ref{eq:TotalRadCorr})). Only after this is done is one allowed to take
$E_{\gamma}^{min}\rightarrow 0$.
The most striking feature of Eq.
(\ref{eq:CFunction}) is that it is independent of $E_{\gamma}^{min}$.
Consequently, it survives in the limit $E_{\gamma}^{min}\rightarrow 0$.
It appears that this procedure was not followed in Ref. \cite{Towner} and,
therefore, equation (44) and Table II in \cite{Towner} must be modified
accordingly. We plot in Fig. \ref{fig:low-thres} the cross section correction
$(\alpha / \pi) g(E_{obs})$, which is for $E_{\gamma}^{min} \rightarrow 0$ independent
of the neutrino energy $E_{\nu}$. Note that it differs in slope compared with its analog
in Table II of  Ref. \cite{Towner}. 

The two aforementioned modifications to the treatment in
Ref. \cite{Towner} allowed us to bring the two cases 
$E_{\gamma}^{min} \rightarrow \infty$ and $E_{\gamma}^{min} \rightarrow 0$ in agreement in
terms of correction to the total cross-section and resolve the discrepancy
mentioned in Ref. \cite{John}.
In either of these extreme cases, by integrating over $E_{obs}$ we obtain
the QED correction to the total cross section
as a function of the neutrino energy $E_{\nu}, ~\delta \sigma^{tot}(E_{\nu})$, displayed
in Fig. \ref{fig:TotalCorection}. 

\subsection{Realistic bremsstrahlung threshold}

The treatment of the more realistic case   
is now straightforward. First and second terms 
(virtual and $E_{\gamma} < E_{\gamma}^{min}$) on the r.h.s of 
Eq. (\ref{eq:TotalRadCorr}) are evaluated by setting 
$E_{\gamma}^{min}=1$ MeV \cite{Hamish} and $E_e=E_{obs}$. 
In the third term
one has to set 
$E_{\gamma}+E_e=E_{obs}+E_{\gamma}^{min}=\rm const$ 
and integrate over $E_e$.

In particular, suppose we write the double differential cross-section 
for the $d+\nu_e \rightarrow p+p+e+\gamma$ as 
$d^2 \sigma_{CC}^{\gamma} / (dE_e dE_{\gamma}) = f(E_e,E_{\gamma})$. 
Then the total cross-section with $E_{\gamma}> E_{\gamma}^{min}$ MeV is:

\begin{eqnarray}
\label{eq:varchange}
\left( \sigma_{CC}^{\gamma} \right)_{tot}= & & {\int}_{m_e}^{E_\nu +\Delta} dE_e 
{\int}_{E_{\gamma}^{min}}^{E_\nu 
+\Delta-E_e} f(E_e,E_{\gamma}) dE_{\gamma}  \nonumber \\
& = & {\int}_{m_e}^{E_\nu +\Delta - E_{\gamma}^{min}} dE_{obs} 
{\int}_{m_e}^{E_{obs}} f(E_e,E_{obs} + E_{\gamma}^{min} - E_e) dE_e ~.
\end{eqnarray}

We simply performed the change of integration variables 
from $(E_e, E_{\gamma})$ to $(E_{obs}, E_e)$ 
in the spirit of Ref. \cite{Towner}. Now we can write in the notation of 
Eq. (\ref{eq:TotalRadCorr}):

\begin{equation}
{\alpha \over \pi}  g_b^{high}(E_{\gamma} \ge E_{\gamma}^{min} )= 
{\int}_{m_e}^{E_{obs}} f(E_e,E_{obs} + E_{\gamma}^{min} - E_e) dE_e ~.
\end{equation}

The result, as expected, is a function of $E_{obs}$ only. 
In order to generalize to the case $E_{e}^{min}>m_e$ 
one has to be careful because change of variables from 
$(E_e, E_{\gamma})$ to $(E_{obs}, E_e)$ becomes less trivial. 
It is possible to show, however, that the following relationship holds:

\begin{eqnarray}
\label{eq:ElThresh}
& & {\alpha \over \pi} 
g_b^{high}(E_{\gamma} \ge E_{\gamma}^{min}, E_{e}^{min}>m_e)= 
{\alpha \over \pi}  g_b^{high}(E_{\gamma} \ge E_{\gamma}^{min}, E_{e}^{min}=m_e) 
\nonumber \\
& + & \int_{m_e}^{E_e^{min}} [ f(E_e,E_{obs} + E_{\gamma}^{min}
- m_e)-f(E_e,E_{obs}+ E_{\gamma}^{min} - E_e) ] dE_e ~,
\end{eqnarray}
where $f(x,y)$ is the function defined before in Eq. (\ref{eq:varchange}). 
Eq. (\ref{eq:ElThresh}) allows one to obtain the correct spectrum, 
Eq. (\ref{eq:TotalRadCorr}), for any electron threshold in terms of the 
ideal case where all electrons are detected. 
We note that it is only the third term in Eq. (\ref{eq:TotalRadCorr}) 
that (implicitly) depends on $E_e^{min}$. 
The impact of the refinement in Eq. (\ref{eq:ElThresh}) is rather small 
for low values of $E_e^{min}$. We evaluated it for $E_e^{min}=1.5$ MeV 
(1 MeV kinetic energy). The effect of the 
second line in Eq. (\ref{eq:ElThresh}) is a 0.03\% modification
of the differential cross section. Consequently, 
we neglect this refinement in our analysis. 

\begin{figure}
\hspace{0.00in}
\begin{center}
\includegraphics[width=4.7in]{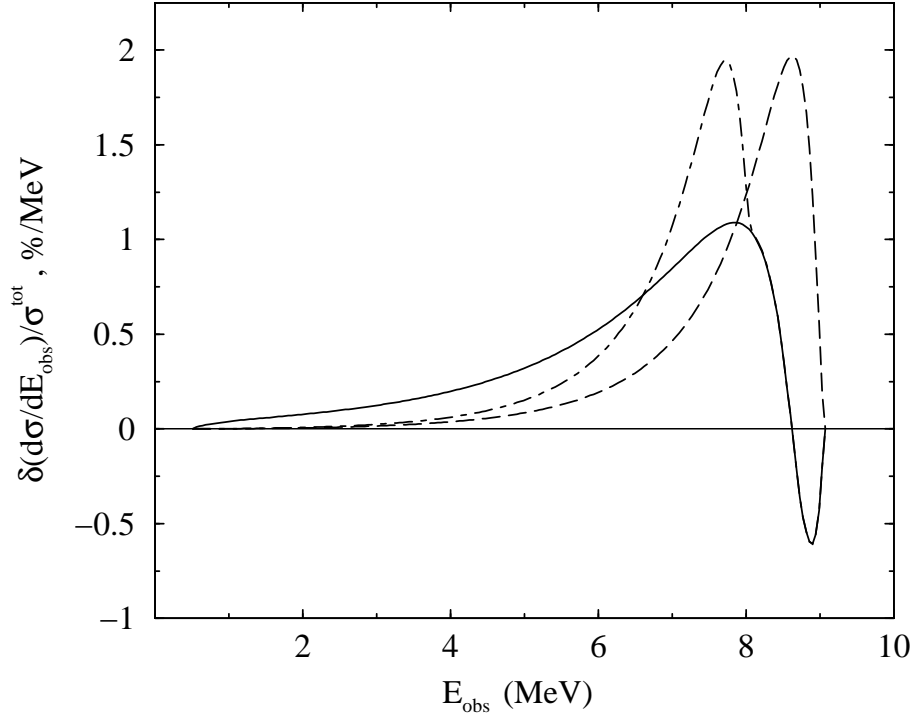}
\end{center}
\caption{\label{fig:differential} Corrections to the differential
cross-sections as a function of the observed energy, normalized to the total
tree-level cross-section. Solid line corresponds to
$E_{\gamma}^{min}\rightarrow \infty$ ($E_{obs}=E_e$), dashed line to
$E_{\gamma}^{min}\rightarrow 0$ ($E_{obs}=E_e+E_{\gamma}$), and
the dot-dashed line is obtained by setting
$E_{obs}=E_e+(E_{\gamma}-1)\theta(E_{\gamma}-1)$. All lines are evaluated
for $E_{\nu}$ = 10 MeV.}
\end{figure}

\begin{figure}
\begin{center}
\includegraphics[width=4.0in]{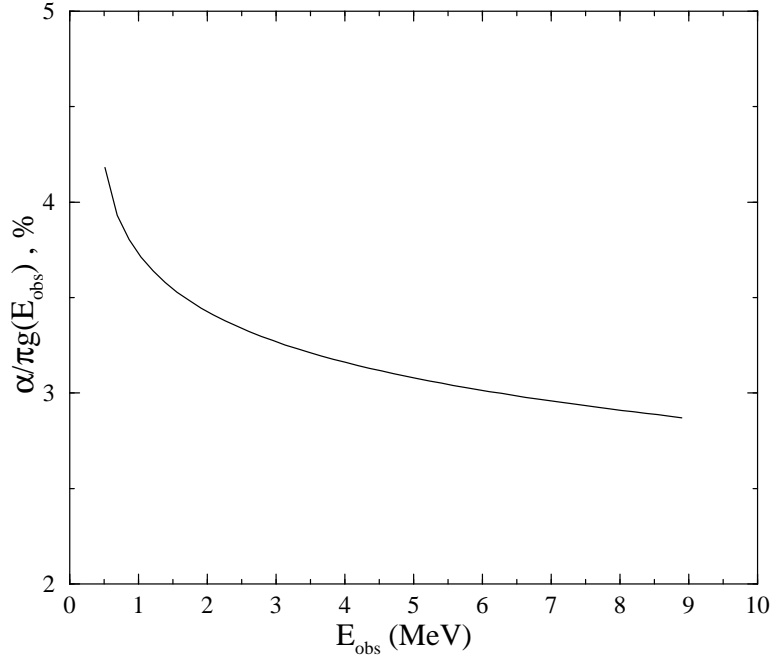}
\end{center}
\caption{Correction to the cross section for $E_{\gamma}^{min} \rightarrow 0$.
Note that in this case the corrections does not depend on $E_{\nu}$.}
\label{fig:low-thres}
\end{figure}

The spectrum for the case C) is shown in Fig. \ref{fig:differential} in the dash-dotted line.
As expected, the upper 1 MeV of that spectrum coincides with the
$E_{\gamma}^{min} \rightarrow \infty$ case. Note that the areas under all three
cases in Fig. \ref{fig:differential} are the same, as they must for consistency.

In Table \ref{tab:spectrum} we provide detailed tabular information on the correction to the
differential cross section for the full range of neutrino energies $E_{\nu}$
and $E_{obs}$ for the case C). 

\section{Folding with the $^8$B spectrum.}

In an actual solar neutrino experiment, like SNO, the recorded quantity is the number of
events with energy $E_{obs}$ (or the total number of events integrated over $E_{obs}$).
This is an integral over the product of the incoming neutrino spectrum and the differential
cross section, i.e.

\begin{equation}
\frac{{\rm d}\sigma}{{\rm d}E_{obs}} = 
\int_0^{\infty} \frac{{\rm d}\sigma (E_{\nu})}{{\rm d}E_{obs}} 
f(E_{\nu}){\rm d}E_{\nu} ~,
\label{eq:folded}
\end{equation}
where $f(E_{\nu})$ is properly normalized incoming neutrino spectrum, possibly modified
by neutrino oscillation. When testing the `null hypothesis', i.e. asking whether neutrinos
oscillate, one takes for the incoming neutrino spectrum simply the shape of the $\nu_e$
spectrum from $^8$B decay \cite{BH} (normalized to unity over the whole range of $E_{\nu}$). 

In Fig. \ref{fig:folded} we show the folded correction to the differential cross section,
Eq. (\ref{eq:folded}) (full line). The case C), i.e. the realistic bremsstrahlung 
detection threshold has been used to produce the full curve. 
For comparison we also show similarly folded tree level
cross section, scaled by a factor 1/40 so that it fits in the same figure
(dashed line). One can see
that the two curves are similar in shape which is basically dictated by the
incoming $^8$B spectrum, but the QED correction is shifted toward smaller
$E_{obs}$, roughly by the value $E_{\gamma}^{min}$ = 1 MeV.

When integrated from the 
threshold used in the SNO analysis, $E_{obs}^{min} - m_e$ = 6.75 MeV,
the full line represents $\sim$ 2\% increase of the total total cross section, therefore
$\sim$ 2\% decrease of the deduced flux, Eq. (\ref{fl:CC}), when the radiative corrections
are properly included. If it would be possible to reduce the threshold to very low
values, the reduction of the flux would be $\sim$ 3\%.

These relative increases of the total cross section obviously differ
somewhat from the values displayed in Fig. \ref{fig:TotalCorection} that were
obtained for monochromatic neutrinos. The difference is, naturally, caused
by the effect of the
shape of the radiative correction to the differential cross section 
in combination with the shape of the $^8$B $\nu_e$ spectrum.
In particular, for the actual SNO $E_{obs}^{min}$ threshold one could have
expected an increase of the cross section (or count rate) due to radiative 
correction of $\sim$ 3\% based on Fig. \ref{fig:TotalCorection} while
the folding with the incoming $^8$B spectrum reduces this value to  $\sim$ 2\%.

\section{Radiative corrections to the NC cross
section}

The NC cross section is governed by the effective four fermion
low-energy Lagrangian
\cite{Mus94}
\begin{equation}
\label{eq:lnuhad}
{\cal
L}^{\nu-had} = - {G_F\over
2\sqrt{2}}{\bar\nu}
\gamma^\mu(1-\gamma_5)\nu
\left[\xi_V^{T=1}V_\mu^{T=1}+\xi_V
^{T=0} V_\mu^{T=0}+
\xi_A^{T=1}A_\mu^{T=1}+\xi_A^{T=0} A_\mu^{T=0}\right]\
\ \ ,
\end{equation}
where
\begin{eqnarray}
V_\mu^{T=1} =
\frac{1}{2}\left[{\bar u}\gamma_\mu u - {\bar d}\gamma_\mu d
\right]&\ \ \
\ &
  V_\mu^{T=0}=\frac{1}{2}\left[{\bar u}\gamma_\mu u + {\bar
d}\gamma_\mu d
\right] \\
A_\mu^{T=1} = \frac{1}{2}\left[{\bar
u}\gamma_\mu\gamma_5 u - {\bar
d}\gamma_\mu
	\gamma_5 d \right]&\ \ \ \
&
 A_\mu^{T=0} =  \frac{1}{2}\left[{\bar u}\gamma_\mu \gamma_5 u +
{\bar
d}\gamma_\mu
  \gamma_5 d \right]\ \ \ ,
\end{eqnarray}
and where
only the effects of up- and down-quarks have been included.
At tree level
in the Standard Model, one has
\begin{eqnarray}
\xi_V^{T=1} =
2(1-2\sin^2\theta_W)& \ \ \ \ &\xi_V^{T=0}= -4\sin^2\theta_W \\
\xi_A^{T=1}
= -2& \ \ \ \ &\xi_A^{T=0}=0 \ \ \ .
\end{eqnarray}
As noted in Ref.
\cite{Towner}, the incident and scattered neutrinos do not
contribute to
the bremsstrahlung cross section at ${\cal O}(G_F^2\alpha)$,
while radiation of real photons from
the participating hadrons is
negligible. Thus, the dominant radiative
corrections involve virtual
exchanges, which modify the $\xi_{V,A}^T$ from their tree-level
values:
\begin{equation}
\xi_{V,A}^T\rightarrow
\xi_{V,A}^T\vert_{tree}(1+R_{V,A}^T)\ \ \ ,
\end{equation}
where the
$R_{V,A}^T$ contain the ${\cal O}(\alpha)$ corrections. Since the
NC
amplitudes are
squared in arriving at the cross section, the total
correction to the NC cross section will go as
twice the relevant
$R_{V,A}^T$. (In the notation of Ref. \cite{Towner},
$R_A^{T=1} = (\alpha/2\pi) g_v^{NC}$.)

As emphasized in Ref. \cite{Towner},
considerable simplification follows
when one considers only the
dominant break-up channel: $^3$S$_1(T=0)\to^1$S$_0(T=1)$. As a $\Delta
T=1$, pure spin-flip
transition, this amplitude is dominated at low-energies by the Gamow-Teller
operator. Magnetic
contributions are of recoil order and, thus, $v/c$ suppressed. Consequently,
we need retain only the
$A_\mu^{T=1}$ term in Eq. (\ref{eq:lnuhad}) and consider only the
correction $R_A^{T=1}$.

The source of corrections to $R_A^{T=1}$ include  corrections to the $W$
and $Z$-boson
propagators (Fig. \ref{fig:feynman-NC}, graph (a)); 
electroweak and QED vertex corrections to the $Z\nu\nu$ and
$Zqq$ couplings (Fig.\ref{fig:feynman-NC}, graph (b)); 
external leg corrections (Fig. \ref{fig:feynman-NC}, graph (c)); 
and  box diagrams involving
the exchange of two $W$'s or two $Z$'s
(Fig. \ref{fig:feynman-NC}, graph (d)). The presence of $W$-boson
propagator corrections arises
when the NC amplitude is normalized to the Fermi constant  $G_F$ 
determined from muon decay.
Only the difference between the gauge boson propagator corrections enters
the $R_{V,A}^T$ in this
case. Note that $Z$-$\gamma$ mixing does not contribute to $R_A^{T=1}$
since the neutrino has no EM
charge and the photon has no axial coupling to quarks at $q^2=0$.
Similarly, one encounters no
$Z\gamma$ box diagrams for neutrino-hadron scattering.

In the analysis of Ref. \cite{Towner}, only the $ZZ$ box contribution was
included, yielding a
correction $R_A^{T=1}\approx 0.002$. Inclusion of all diagrams, however,
produces a substantially
larger correction. From the up-dated tabulation of effective $\nu$-$q$
couplings given in Ref.
\cite{PDG}, we obtain
\begin{equation}
R_A^{T=1} = \rho^{NC}_{\nu N} +
\lambda_{dR}-\lambda_{uR}+\lambda_{uL}-\lambda_{dL} -1
\approx 0.0077\  ,
\end{equation}
where we have followed the notation of Ref. \cite{PDG}. In particular, the
$WW$ box graph
contributes roughly 80\% of the total:
\begin{equation}
R_A^{T=1}(WW-{box}) = {5\alpha\over 8\pi\sin^2\theta_W}\approx
0.0063\  .
\end{equation}
The net effect of the total correction is
therefore $g_v^{NC}$ = 6.63, i.e., $\sim 1.5\%$ increase in the NC
cross section,
as compared to the $0.4\%$ increase quoted in Ref. \cite{Towner}.

\section{Conclusions}

\begin{figure}
\hspace{0.00in}
\begin{center}
\includegraphics[width=4.0in]{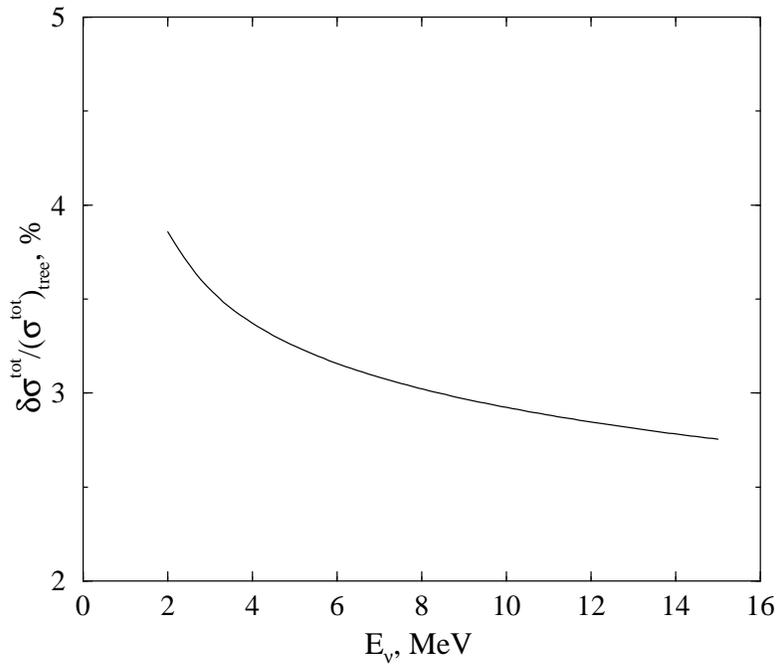}
\end{center}
\caption{\label{fig:TotalCorection} Radiative corrections to the
CC total cross section as a function of neutrino energy.
}
\end{figure}

\begin{figure}
\hspace{0.00in}
\begin{center}
\includegraphics[width=4.5in]{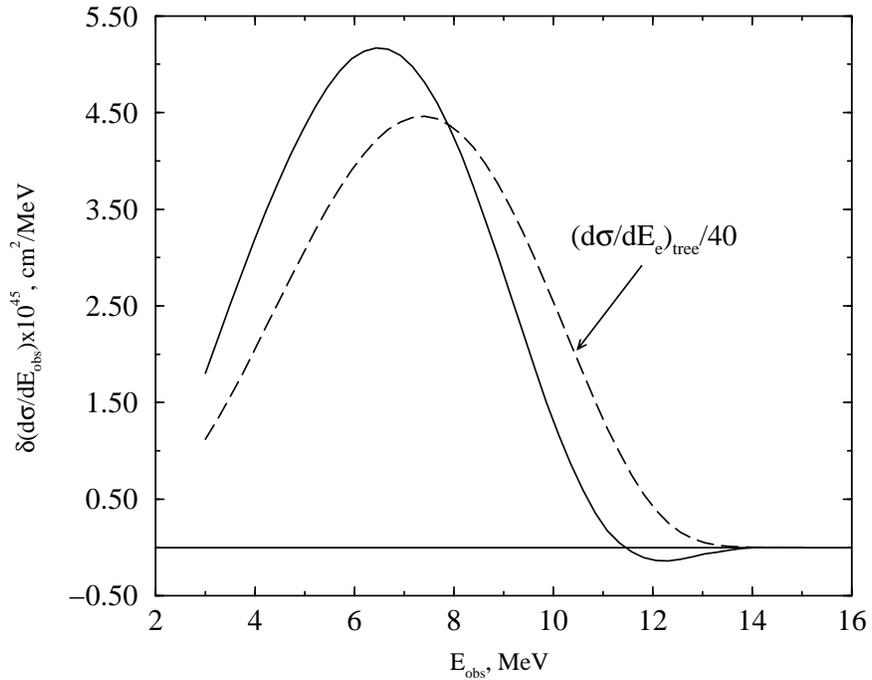}
\end{center}
\caption{\label{fig:folded} Radiative corrections to the
total CC cross section folded with the incoming $^8$B $\nu_e$ spectrum
as a function of the detected energy $E_{obs}$ (full line)
Also shown is similarly folded tree level CC reaction cross section,
scaled by a factor of 1/40 (dashed line).}
\end{figure}

The radiative correction of order $\alpha$ for the charged and neutral
current  neutrino-deuterium disintegration and energies relevant to the SNO
experiment are consistently evaluated. For the CC reaction the contribution
of the virtual $\gamma$ and $Z$ exchange is divided into high and low momentum
parts, and the dependence on the corresponding scale $\mu \sim$ 1 GeV separating
the two regimes is discussed in detail. For the bremsstrahlung emission
we discuss the important role of the bremsstrahlung detection threshold
$E_{\gamma}^{min}$. In particular, we consider the two extreme cases
$E_{\gamma}^{min} \rightarrow \infty$ and $E_{\gamma}^{min} \rightarrow 0$
as well as a more realistic intermediate case. We show that our treatment,
unlike Ref. \cite{Towner}, gives consistent, i.e. $E_{\gamma}^{min}$-independent, 
correction
to the total cross section, shown in Fig. \ref{fig:TotalCorection}.
This correction, slowly decreasing with increasing neutrino energy $E_{\nu}$,
amounts to $\sim$ 4\% at low energies and  $\sim$ 3\% at the end
of the $^8$B spectrum. 

The magnitude of this correction is in accord
with the correction to the inverse neutron beta decay,
$\bar{\nu}_e + p \rightarrow n + e^+$ evaluated in Refs. \cite{vogel,fayans}
and with the correction for the $pp$ fusion reaction evaluated
in Ref. \cite{pp}. Note that in these references only the `outer radiative
corrections', i.e. only the low momentum part of the virtual photon exchange
was considered. The high momentum part, which is independent of the incoming
or outgoing lepton energies, and which is universal for all semileptonic
weak reactions involving $d \leftrightarrow u$ quark transformation, amounts
to $\sim$ 2.4\% \cite{Towner2} and should be added to the results quoted in 
Refs. \cite{vogel,fayans,pp}. 
   
We identify the origin of the inconsistency of the treatment of Ref.  \cite{Towner}: 
(a) neglect of a strong momentum-dependence in the Gamow-Teller
$^3$S$_1\to^1$S$_0$
matrix element, which affects the case 
of $E_{\gamma}^{min} \rightarrow \infty$, 
and (b) improper ordering of limits involving
$E_\gamma^{min}$ and the infrared regulator,
which affects the case of $E_{\gamma}^{min} \rightarrow 0$.
For the more realistic choice of $E_{\gamma}^{min}$ we provide a detailed
evaluation of the correction to the differential cross section.

We also discuss the effect of folding the cross section with the (unobserved
directly) spectrum of the $^8$B decay. We conclude that for the realistic 
choice of $E_{\gamma}^{min}$ and for the electron detection threshold of the
SNO collaboration, the solar $^8$B $\nu_e$ flux deduced neglecting the radiative
correction would be overestimated by $\sim$2 \%. 

\begin{figure}
\includegraphics[width=4.55in]{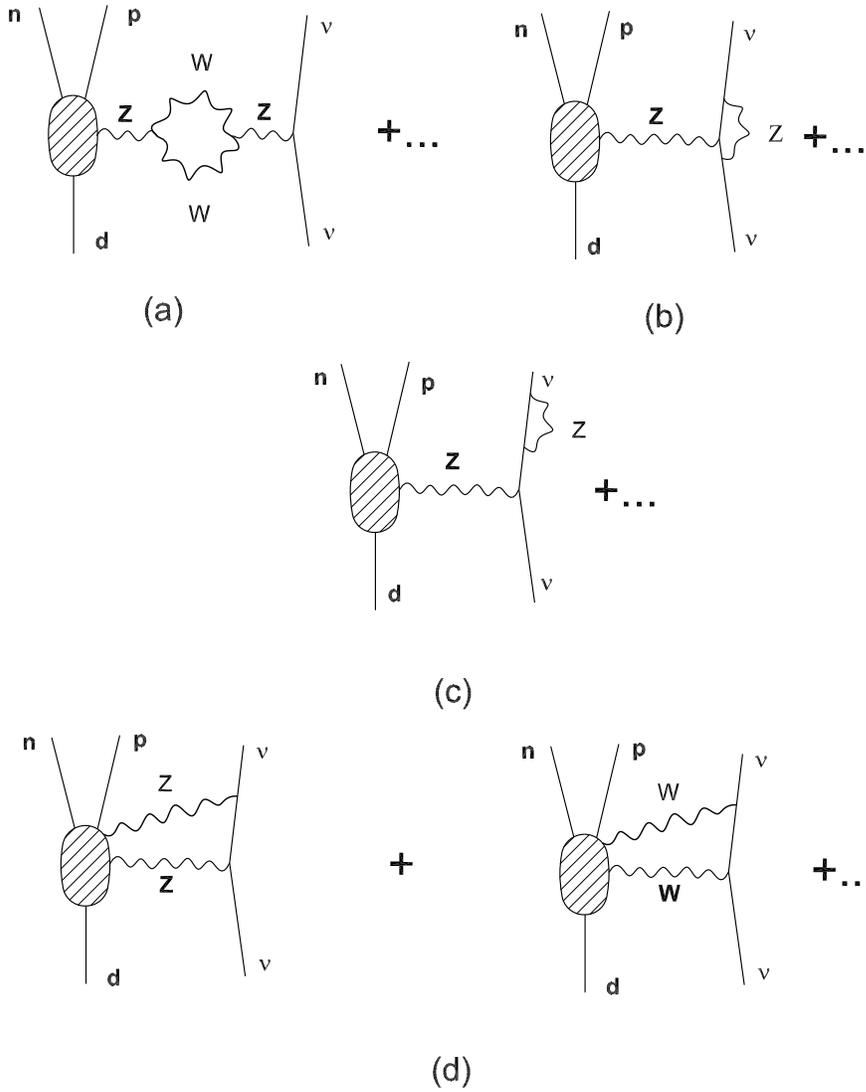}
\vspace{0.3cm}
\caption{Feynman graphs relevant for the radiative correction
to the NC cross section (see text for explanation).}
\label{fig:feynman-NC}
\end{figure}

Next we consider the effect of radiative corrections to the neutral current
deuteron disintegration, so far not analyzed by the SNO collaboration. 
In that case the radiative corrections, associated with the Feynman graphs 
in Fig. \ref{fig:feynman-NC}, are dominated  by the virtual $Z$ and $W$
exchange, in particular be the box graph in Fig. \ref{fig:feynman-NC} (d).
The corresponding neutrino energy independent correction to the NC
total cross section is $\sim$ 1.5\%. 

We provide in the Appendix a set of formulae relevant for the case of an arbitrary
bremsstrahlung threshold $E_{\gamma}^{min}$. These formulae allow one to evaluate
the CC reaction differential cross section in terms of the electron energy
$E_e$, the photon energy $E_{\gamma}$, and the angle between the momenta of the
electron and photon.

\begin{acknowledgments} We would like to thank John Beacom, Art McDonald, and Hamish
Robertson for valuable discussions. This work was supported in part by the 
NSF Grant No. PHY-0071856
and by the U. S. Department of Energy Grants No. DE-FG03-88ER40397 and
DE-FG02-00ER4146.
\end{acknowledgments}

\newpage

\appendix{{\bf Appendix}}

Here we provide a recipe for obtaining radiative 
correction to differential cross-section for the 
reaction $d+\nu_e \rightarrow e+p+p$. The  prescription 
is infrared finite and allows arbitrary values of cutoffs for detection 
of electrons and photons.

Unlike the discussion in the main text here we do 
not choose any particular model for what an experiment 
can detect. The only assumption that we make is that 
the bremsstrahlung photons cannot be seen below certain 
energy $E_{\gamma}^{min}$. Therefore, contributions from 
all photons with energies below this cutoff are added 
and the only thing that is left for detection is the electron energy.

We make no assumptions as to how photons with 
$E_\gamma>E_{\gamma}^{min}$ are recorded. 
For their contribution we provide triple differential 
cross-section that depends on electron energy, 
photon energy and angle between the direction 
of the electron and emitted photon. This expression 
can be incorporated in the detector-specific simulation 
software for appropriate analysis.

We combine the contributions from photons with 
$E_{\gamma}<E_{\gamma}^{min}$ with virtual photon 
and Z exchanges to get infrared finite answer 
(first two terms in Eq. (\ref{eq:TotalRadCorr})). The result is:

\begin{eqnarray}
\label{eq:TotLowEn}
{\left( d \sigma (E_e, E_\nu) \over {dE_e} \right)}_{(E_{\gamma} <
E_{\gamma}^{min})}&=&{\left( d \sigma (E_e, E_\nu) \over {dE_e}
\right)}_{tree} {\alpha \over \pi}  \biggl\{   2~{\rm ln}\left(
{E_{max}\over m_e}\right)\left[\frac{1}{2 \beta}  \ln  \left( \frac{1 +
\beta}{1 - \beta} \right) -1\right]
\nonumber \\
& & +I_1(E_{max},E_\nu)+I_2(E_{max},E_\nu)+{\cal C}(\beta)+{\cal
A'}(\beta)-{3 \over 8} \nonumber \\
& & +g_v^{p \gtrsim \mu} + \frac{3}{2} {\rm ln} \left( \frac{\mu}{M_p}
\right)
+ 3{\bar Q} {\rm ln} \left( \frac{\mu}{M_A} \right)\nonumber \\
& & +\int_0^{E_{max}} \left({{|I(E_{\nu} + \Delta - E_{obs} -
E_{\gamma})|^2}
\over {|I(E_{\nu} + \Delta - E_{obs})|^2}} - 1\right) 
\frac{{\rm d} g_b(E_{obs}, E_{\nu}, E_{\gamma})}{{\rm d}
E_{\gamma}}dE_{\gamma} \biggr\}    ~ \nonumber \\ \\
E_{max}&=&{\rm Min} \left[ E_{\gamma}^{min}, E_\nu+\Delta-E_e \right] ~. 
\end{eqnarray}

where $\cal C(\beta)$ is defined in Eq. (\ref{eq:CFunction}), 
$g_v^{p \gtrsim \mu}$ is taken from Eq. (\ref{eq:gvshort}), 
and ${\cal A'}(\beta)$, $I_1(E_{max},E_\nu)$, 
and $I_2(E_{max},E_\nu)$ are defined as follows (see \cite{Towner}):

\begin{eqnarray}
\label{TotLowEn-Func}
{\cal A'}(\beta)&=&\frac{1}{2} \beta ~ {\rm ln} \left( \frac{1 + \beta}{1 - \beta}
\right) -1 + \frac{3}{2} ~ {\rm ln}  \left( \frac{ M_p } { m_e } \right)
- \frac{1}{ \beta }
\left[ \frac{1}{2} {\rm ln} \left( \frac{1 + \beta}{1 - \beta} \right)
\right]^2
+ \frac{1}{ \beta } L \left( \frac{ 2 \beta }{ 1 + \beta } \right) \nonumber \\
I_1(E_{max},E_\nu)&=& -{1\over{\beta ~ E_e^2}} ~ 
{\rm ln}\left( \frac{1 + \beta}{1 - \beta} \right)
{ {(E_\nu + \Delta -E_e)^2} \over 15} \nonumber \\
& & \times \left[\left( 5-3\left( 1-{E_{max}\over 
{E_\nu +\Delta-E_e}}\right)\right)
\left( 1-{E_{max}\over {E_\nu +\Delta-E_e}}\right)^{3/2}-2\right]\nonumber \\
I_2(E_{max},E_\nu)&=& 2 \left[\frac{1}{2 \beta}  
\ln  \left( \frac{1 + \beta}{1 - \beta} \right) -1\right] \nonumber \\
& & \times \int_0^{E_{max}}\left[ 
\left(  1-{Q \over {E_\nu +\Delta-E_e}} \right)^{1/2} 
\left (  1+{Q \over E_e}\right) -1\right]{dQ \over Q}~.
\end{eqnarray}

For $E_{\gamma}>E_{\gamma}^{min}$ we write 
the triple differential cross-section

\begin{eqnarray}
\label{eq:triple}
{\left( d \sigma (E_e, E_\nu) \over 
{dE_e~dE_{\gamma}~dx} \right)}_{(E_{\gamma} >  
E_{\gamma}^{min})}&=& {\alpha \over {\pi} } 
\frac{G_F^2}{\pi} V_{ud}^2 g_A^2 M_p \beta(E_e) E_e^2 \nonumber \\
& & \times (M_p(E_\nu +\Delta -E_e-E_{\gamma}))^{1/2} 
|I(E_\nu +\Delta -E_e-E_{\gamma})|^2 \nonumber \\
& &\times E_{\gamma} \left[ \frac{1}
{E_e^2 (1 - \beta ~ x)} +
\beta^2 \frac{E_e + E_{\gamma}}{E_e~E_{\gamma}^2} \frac{1 -  x^2}
{(1 - \beta~x)^2} \right]
\end{eqnarray}
where $x$ is cosine of the angle between 
photon and electron momenta. We have integrated 
over the corresponding azimuthal angle.

\begin{table}[htb]
\caption{Values of the correction to the differential cross-section,  
$\delta \left( {{d\sigma(E_\nu,E_{obs})} \over {dE_{obs}}}\right) / \sigma^{tot}(E_\nu)$,
normalized to the total tree-level cross-section,  in \%/MeV.
The neutrino energy $E_\nu$ in MeV labels the columns,
while the total energy observed in the detector, $E_{obs}$, 
in the form $E_\nu+\Delta-E_{obs}$ also in MeV,
labels the lines. The dash-dotted curve in Fig. \ref{fig:differential} 
corresponds to the column with $E_\nu$=10 MeV.}
\label{tab:spectrum}
\vspace{0.3cm}
\begin{tabular}{ r | r r r r r r r r r r r r r r }

$E_\nu+\Delta-E_{obs} $&$E_{\nu}$: 2& 3& 4& 5& 6& 7& 8& 9& 10& 11& 12& 13& 14& 15\\
\hline
0.1       &6.15 &1.44 &0.59 &0.21 &-0.02&-0.18&-0.31&-0.40&-0.48&-0.55&-0.62&-0.67&-0.72&-0.79\\
0.2       &10.64&3.22 &1.60 &0.83 &0.36 &0.02 &-0.23&-0.43&-0.60&-0.75&-0.88&-0.99&-1.10&-1.25\\
0.3       &10.57&4.17 &2.31 &1.37 &0.77 &0.35 &0.02 &-0.24&-0.47&-0.66&-0.82&-0.98&-1.11&-1.26\\
0.4       &7.88 &4.45 &2.73 &1.78 &1.16 &0.71 &0.35 &0.07 &-0.17&-0.38&-0.57&-0.73&-0.88&-1.02\\
0.5       &4.00 &4.32 &2.92 &2.06 &1.47 &1.03 &0.69 &0.41 &0.17 &-0.04&-0.23&-0.39&-0.54&-0.67\\
0.75      &     &3.16 &2.71 &2.22 &1.82 &1.49 &1.22 &1.00 &0.80 &0.63 &0.48 &0.34 &0.21 &0.10\\  
1.        &     &1.84 &2.10 &1.93 &1.72 &1.52 &1.34 &1.18 &1.04 &0.91 &0.80 &0.69 &0.60 &0.51\\ 
1.25      &     &0.81 &1.66 &1.86 &1.91 &1.92 &1.91 &1.89 &1.88 &1.86 &1.85 &1.83 &1.82 &1.81\\ 
1.5       &     &0.16 &1.08 &1.42 &1.59 &1.69 &1.75 &1.80 &1.84 &1.88 &1.91 &1.94 &1.96 &1.98\\ 
1.75      &     &     &0.54 &0.97 &1.16 &1.29 &1.37 &1.44 &1.49 &1.54 &1.58 &1.61 &1.65 &1.68\\ 
2.        &     &     &0.29 &0.63 &0.82 &0.94 &1.03 &1.10 &1.15 &1.20 &1.24 &1.27 &1.31 &1.34\\ 
2.25      &     &     &0.13 &0.40 &0.57 &0.69 &0.77 &0.83 &0.89 &0.93 &0.96 &1.00 &1.03 &1.05\\
2.5       &     &     &0.03 &0.25 &0.40 &0.50 &0.58 &0.64 &0.68 &0.72 &0.75 &0.78 &0.81 &0.83\\ 
2.75      &     &     &     &0.13 &0.27 &0.37 &0.43 &0.49 &0.53 &0.56 &0.59 &0.62 &0.64 &0.66\\ 
3.        &     &     &     &0.07 &0.18 &0.27 &0.33 &0.37 &0.41 &0.44 &0.47 &0.49 &0.51 &0.53\\ 
3.5       &     &     &     &0.01 &0.08 &0.14 &0.19 &0.23 &0.26 &0.28 &0.30 &0.32 &0.34 &0.35\\ 
4.        &     &     &     &     &0.02 &0.07 &0.11 &0.14 &0.16 &0.18 &0.20 &0.22 &0.23 &0.24\\ 
4.5       &     &     &     &     &     &0.03 &0.06 &0.08 &0.10 &0.12 &0.14 &0.15 &0.16 &0.17\\ 
5.        &     &     &     &     &     &0.01 &0.03 &0.05 &0.07 &0.08 &0.09 &0.10 &0.11 &0.12\\ 
\end{tabular}

\end{table}


\newpage

\end{document}